\documentclass[useAMS,usenatbib]{mn2e} 
\usepackage{aas_macros}
\usepackage{graphicx}
\usepackage{deluxetable}

\title{3C273 variability at 7 mm: Evidences of shocks and precession in the jet.}

\author[P.P.B.Beaklini and Z.Abraham]
{Pedro Paulo B. Beaklini$^{1}$
\newauthor
Zulema Abraham$^1$\\
$^1$Instituto de Astronomia,
Geof\'{\i}sica e Ci\^{e}ncias Atmosf\'{e}ricas, Universidade de S\~{a}o Paulo.\\
Rua do Mat\~{a}o 1226, 05508-090, S\~{a}o Paulo/SP, Brazil.}
\date{Released 2011 Xxxxx XX}

\begin{document}

\maketitle

\begin{abstract}

We report 4 years of observations of 3C273 at 7 mm obtained with the Itapetinga Radiotelescope, in Brazil, between 2009 and 2013.  We detected a flare  in 2010 March, when the flux density increased by 50\% and reached 35 Jy. After the flare, the flux density started to decrease and reached values lower than 10 Jy. We suggest that  the 7 mm flare is the radio counterpart of the  $\gamma$-ray flare observed by Fermi/LAT in  2009 September, in which the flux density at high energies  reached a factor of fifty of its average value.  A delay of 170 days between the radio and $\gamma$-ray flares was revealed using the Discrete Correlation Function (DCF) that can be interpreted in the context of a shock model, in which each flare corresponds to the formation of a compact superluminal component that expands and becomes optically thin at radio frequencies at latter epochs. The difference in flare intensity between frequencies and at a different times, is explained as a consequence of an increase in the Doppler factor $\delta$, as predicted by the 16 year precession model proposed by Abraham \& Romero, which has a large effect on boosting  at high frequencies while does not affect too much the observed optically thick  radio emission. We discuss other observable effects of the variation in $\delta$, as the increase in the formation rate of superluminal components, the variations in the time delay between flares and the periodic behaviour of the radio light curve that we found  compatible with changes in the Doppler factor.

\end{abstract}

\begin{keywords}
(galaxies:) quasars: individual: 3C273
\end{keywords}

\section{Introduction}

Even though the launch of the Fermi Gamma-ray Space Telescope resulted in an unprecedented  amount of data, which complemented the already known lower frequency spectral energy distribution (SED) of active galactic nuclei (AGNs), the actual emission process at each frequency is still under debate. There is no doubt that the radio emission has synchrotron origin, but the high energy X- and $\gamma$-rays can be attributed to different processes, like inverse Compton up scattering of low frequency photons, either of synchrotron origin (Synchrotron Self Compton, SSC), or external (External Compton, EC), or even hadronic processes initiated by relativistic protons \citep[see review by][]{bot10}.

Besides a quiescent or slowly varying emission, blazars  present short duration high energy $\gamma$-ray flares with intensities that can differ in several orders of magnitude, even for flares in the same object. Similar but longer lasting flares are observed at infrared and radio frequencies, the latter associated with the appearance of relativistically beamed superluminal components in the parsec scale jets \citep{jor01}. As the radio emission, the $\gamma$-ray flux must be also relativistically beamed, to account for the short variability timescales and for the small optical depth for pair production \citep{mat93,weh98}. 

The superluminal components do not have the same apparent velocity and position angle in the plane of the sky \citep[eg.][]{cot79}, and in the case of 3C273, the systematic variation of these quantities were interpreted as due to jet precession, assuming ballistic motion for the components \citep{abr99}. The period detected was 16 years, which together with the black hole mass, suggested that the Bardeen-Peterson effect could be the origin of this precession \citep{cap04}. 

Considering only single dish observations, the radio emission of 3C273 was extensively monitored at different frequencies \citep{tur99,sol08}. This long time  coverage (almost 40 years) allowed statistical studies that revealed the existence of several periodicities, including  8 and 32 years \citep{fan07,zha10,vol13}. 

\begin{figure*}
  \includegraphics[width=18 cm]{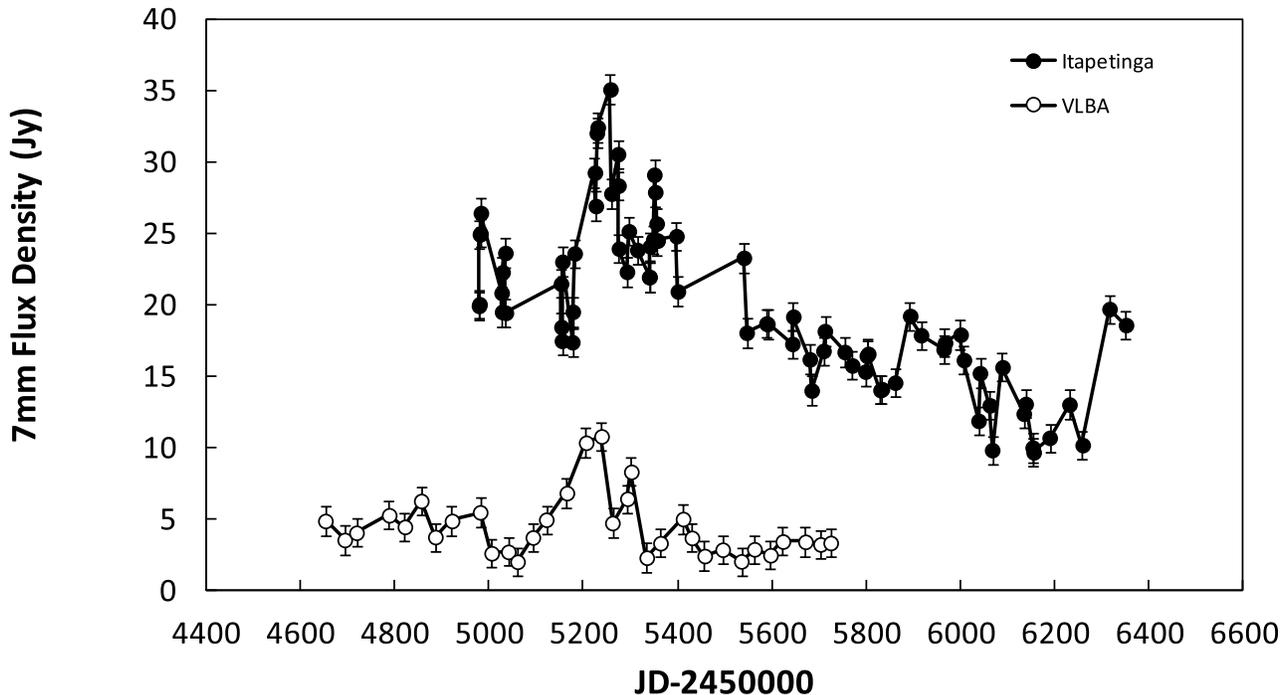}
  \caption{The 7 mm single dish light curve of 3C273 (this work) and the flux densities of the obtained with the VLBA by \citet{mar12}.}
    \label{lc}
\end{figure*}

The formation of the superluminal components was attributed to particle acceleration  in shocks propagating along the relativistic jet \citep{mar85,hu85,spa01,mar10,hug11},  and their temporal  evolution  at different frequencies was extensively explored, \citep[eg.][]{tur00,bot10}. Since initially the shocked region is optically thick at radio frequencies, a delay  between the maxima in the light curves at different frequencies is expected.
 
The detection of time delays is not easy, because the association of the flares at different frequencies is not unique, and even in the same source, each flare could have a different time delay. Before the Fermi era, a time delay of hundred days was detected in 3C273,  between  radio and infrared flares, with the infrared flare coming first \citep{cle83,rob83,bot88,ste94,ste98}. 

Long time delays (1 to 8 months) between radio and $\gamma$-ray emission in Fermi blazars were measured comparing  the light curves of 186 sources from the MOJAVE program \citep{lis09} with the Fermi results \citep{pus10}. In 3C279, a time delay of 6 months between the light curves  was  detected  \citep{cha10}. However, up to now there was no mention of time delay between radio and $\gamma$-rays in the literature for 3C273, although the relation between these frequencies was discussed in several occasions \citep{jor01,jor12,mar12}.

In this paper we present 7 mm (43 GHz) single dish observations of 3C273 and associated the radio flare detected in 2010 March with the $\gamma$-ray flare observed by Fermi/LAT in 2009 September (Section 2). We discussed the time delay in terms of a shock model an interpreted the high intensity of the $\gamma$-ray flare as a consequence of an increase in the Doppler factor as predicted by the precession model of \citet{abr99} (Section 3).   
 At last, in Section 4 we presented our conclusions.

\section{Observations and Results}
\label{Observations and Results}

The observations  of 3C273 at 7mm were made  with the 13.7 m radome enclosed Itapetinga radio telescope, between 2009 and 2013. At this wavelength, the antenna half power beam width (HPBW) is about 2.4 arcmin and the radome transmission 0.68. The receiver, a room temperature K-band mixer, has a 1-GHz double side band (d.s.b) and a noise temperature of about 700 K. The calibration was made with a known temperature noise source and a room temperature load, which automatically corrects for atmospheric attenuation and  radome absorption \citep{abr92}. The HII region SgrB2 Main was used as a primary flux calibrator. The method of observation consisted of scans in elevation and azimuth with an amplitude of 30 arcmin and 20s duration. The scans in each direction were averaged, a baseline was subtracted to eliminate the contribution of the atmosphere, and a Gaussian with the 7 mm HPBW  was fitted to the remaining data to obtain the flux density; its central position  was used to check the pointing accuracy.

\begin{table*}

 {\small
 \caption{Days of Observations and flux density values}
 \hfill{} 
 \begin{tabular}{ c c c c c c c c }
\hline Date & JD-2450000 & Flux Density & Error & Date & JD-2450000 & Flux Density & Error\\
 & & (Jy) & (Jy) &  & & (Jy) & (Jy) \\ 
\hline 27/05/2009 & 4978 & 19.94 & 0.99 & 27/01/2011 & 5588 & 18.70 & 0.92 \\
28/05/2009 & 4979 & 20.05 & 1.06 & 29/01/2011 & 5590 & 18.66 & 0.80 \\
29/05/2009 & 4980 & 24.95 & 0.92 & 23/03/2011 & 5643 & 17.27 & 0.83 \\
30/05/2009 & 4981 & 25.02 & 0.84 & 25/03/2011 & 5645 & 19.18 & 0.96 \\
31/05/2009 & 4982 & 26.45 & 1.01 & 29/04/2011 & 5680 & 16.19 & 1.10 \\
14/07/2009 & 5026 & 20.85 & 0.95 & 03/05/2011 & 5684 & 14.00 & 1.11 \\
15/07/2009 & 5027 & 19.51 & 0.80 & 28/05/2011 & 5709 & 16.78 & 0.87 \\
16/07/2009 & 5028 & 22.30 & 1.07 & 31/05/2011 & 5712 & 18.17 & 1.00 \\
21/07/2009 & 5034 & 23.65 & 1.11 & 12/07/2011 & 5754 & 16.70 & 0.75 \\
22/07/2009 & 5035 & 19.45 & 0.69 & 28/07/2011 & 5770 & 15.76 & 0.62 \\
17/11/2009 & 5152 & 21.49 & 1.35 & 25/08/2011 & 5798 & 15.33 & 0.60 \\
18/11/2009 & 5153 & 18.45 & 1.00 & 28/08/2011 & 5801 & 16.44 & 0.63 \\
19/11/2009 & 5154 & 17.49 & 0.94 & 31/08/2011 & 5804 & 16.57 & 0.64 \\
20/11/2009 & 5155 & 23.03 & 1.16 & 26/09/2011 & 5830 & 14.03 & 0.64 \\
11/12/2009 & 5176 & 17.39 & 0.91 & 29/09/2011 & 5833 & 14.10 & 0.64 \\
12/12/2009 & 5177 & 19.52 & 1.08 & 27/10/2011 & 5861 & 14.56 & 1.00 \\
16/12/2009 & 5181 & 23.61 & 1.59 & 28/11/2011 & 5893 & 19.23 & 1.11 \\
28/01/2010 & 5224 & 29.28 & 2.33 & 21/12/2011 & 5916 & 17.88 & 1.50 \\
30/01/2010 & 5226 & 26.94 & 1.85 & 07/02/2012 & 5964 & 16.89 & 0.80 \\
01/02/2010 & 5228 & 32.05 & 2.34 & 10/02/2012 & 5967 & 17.35 & 0.89 \\
03/02/2010 & 5230 & 32.46 & 1.65 & 14/03/2012 & 6000 & 17.92 & 1.29 \\
02/03/2010 & 5257 & 35.11 & 2.92 & 21/03/2012 & 6007 & 16.14 & 0.90 \\
04/03/2010 & 5259 & 27.81 & 1.19 & 21/04/2012 & 6038 & 11.88 & 0.92 \\
18/03/2010 & 5273 & 30.57 & 1.16 & 24/04/2012 & 6041 & 15.22 & 1.70 \\
19/03/2010 & 5274 & 28.37 & 1.17 & 15/05/2012 & 6062 & 12.96 & 0.89 \\
20/03/2010 & 5275 & 23.95 & 1.85 & 20/05/2012 & 6067 & 9.83 & 0.74 \\
06/04/2010 & 5292 & 22.32 & 2.01 & 11/06/2012 & 6089 & 15.63 & 1.47 \\
10/04/2010 & 5296 & 25.17 & 0.85 & 26/07/2012 & 6134 & 12.38 & 0.87 \\
28/04/2010 & 5314 & 23.86 & 0.84 & 30/07/2012 & 6138 & 13.06 & 0.64 \\
30/04/2010 & 5339 & 21.95 & 0.79 & 14/08/2012 & 6153 & 10.00 & 0.59 \\
13/05/2010 & 5340 & 21.96 & 0.76 & 16/08/2012 & 6155 & 9.67 & 0.60 \\
15/05/2010 & 5341 & 24.11 & 1.25 & 19/09/2012 & 6189 & 10.69 & 1.34 \\
01/06/2010 & 5348 & 24.59 & 1.02 & 31/10/2012 & 6231 & 13.02 & 1.97 \\
03/06/2010 & 5350 & 29.13 & 0.98 & 28/11/2012 & 6259 & 10.18 & 1.68 \\
05/06/2010 & 5352 & 27.91 & 1.09 & 24/01/2013 & 6316 & 19.72 & 2.10 \\
08/06/2010 & 5355 & 25.70 & 0.88 & 28/02/2013 & 6351 & 18.59 & 1.18 \\
10/06/2010 & 5357 & 24.52 & 1.06 & 03/04/2013 & 6385 & 17.16 & 1.40 \\
20/07/2010 & 5397 & 24.82 & 0.98 & 07/04/2013 & 6389 & 17.36 & 1.39 \\
23/07/2010 & 5400 & 20.95 & 0.72 & 11/04/2013 & 6393 & 18.11 & 0.79 \\
10/12/2010 & 5540 & 23.30 & 0.93 & 10/05/2013 & 6421 & 17.62 & 0.74 \\
16/12/2010 & 5546 & 18.05 & 1.50 & 13/05/2013 & 6425 & 16.90 & 0.95 \\
\hline
\end{tabular}}
\hfill{}
\label{T1}
\end{table*}

The 7 mm data, obtained between 2009 May and 2013 April, are presented in Table 1 and in the upper part of Fig. \ref{lc}. In the first epoch of our monitoring, between 2009 and 2010, we detected a series of flares, the strongest with maximum flux density of  $35.1 \pm 2.9$ Jy on  2010 March 2; in the second epoch, between 2011 and 2013, we detected a continuous decrease in the flux density, reaching the minimum of $9.8 \pm 0.7$ Jy on 2012 March 20. After this minimum, the flux density increased to the  value that it had before the 2010 flare.

\begin{figure}
  \includegraphics[width=\columnwidth]{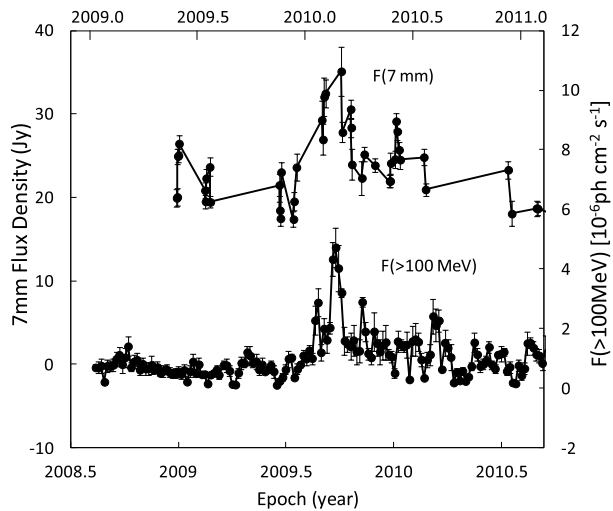}
  \caption{7 mm  and $\gamma$-ray light  curves of 3C273, with time given in the upper and lower axis, respectively. The origin of the radio time axis was shifted by 170 days, to show the similarity between the flares at both frequencies.}
  \label{roi_fermi}
\end{figure}

In total, our light curve has  82 days of observations, with two gaps of about 100 days, between July and November of 2009, and between July and December of 2010. Although the first gap  coincided with the strong $\gamma$-ray flares reported by \citet{abd10} in 2009 September and therefore, for which we do not have simultaneous 7 mm data, a large flare in  radio was observed 6 months later, in 2010 March. The pattern of this radio flare is similar to that of the  2009 flare at $\gamma$-rays, as can be seen in Fig. \ref{roi_fermi}, where we present part our 7 mm light curve (upper part)  and the $\gamma$-ray light curve (lower part) from the Fermi Space telescope\footnote{$http: //fermi.gsfc.nasa.gov/ssc/data/access/lat/msl\_lc/$} binned in 5 days intervals, with the time axis  displaced by 150 days relative to the 7 mm time axis, so that the flares at the two different wavelengths  became aligned. 

Since the gap in our light curve does not allow us, without any other information, to affirm that the 2010 March radio flare is correlated with the 2009 September $\gamma$-ray  flare, we  analysed the 7 mm VLBA  light curve of the compact core \citep{jor12,mar12}  shown in the lower part of Fig. \ref{lc}\footnote{\citet{mar12} reported the flux density of 3C273 relative to the peak value of the light curve. To calculated the light curve in $Jy$, we used the peak value presented in the website of Boston University gamma-ray blazar monitoring program: $http://www.bu.edu/blazars/VLBAproject.html$}. 
The VLBA flux density does not show any increase in intensity at the epoch of the $\gamma$-ray flare, instead of that,  is was lower than at other epochs, while  it reached its maximum value in 2010 March, at the same time as the Itapetinga light curve. 
We must emphasize the similarity between the two light curves, indicating that the 7 mm single dish observations reflect mainly the core variability. Furthermore, the VLBI images reported by \citet{jor12} do not present any strong variation in the flux density of the  superluminal components at the epoch of the gamma ray flare, also in agreement with our observations.

\begin{figure}
  \includegraphics[width=\columnwidth]{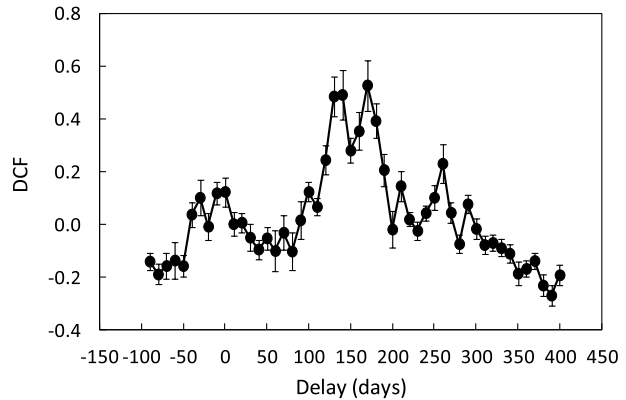}
  \caption{Results of the Discreted Correlation Function: the upper graph corresponds to the correlation of the real radio and $\gamma$-ray data, the lower corresponds to the correlation of simulated radio data with the real $\gamma$-ray light curve.}
  \label{dcf}
\end{figure}

Finally, we  used the DCF (Discrete Correlation Function) to verify statistically the correlation between the 7 mm and $\gamma$-ray flares and to compute the correct time delay.  The  DCF is a simple test which calculates the correlation between two light curves without interpolating or creating data; the maxima indicate the time delays \citep{ede88}. To calculated the DCF, we used the Itapetinga (46 points) and the Fermi light curve (758 points) shown binned in Fig. \ref{roi_fermi}, because the rest of the light curves are not necessarily correlated.
In Fig. \ref{dcf} we presented the DCF that  shows a wide maximum at delays between 120 and 170 days, meaning that the radio flare occurred after the $\gamma$-ray flare.

\section{Discussion}
\subsection{Flare intensities and the beaming factor}

The first flare detected in 3C273 and monitored at different wavelengths (between $100 \mu$m and $8.9$ mm) was reported by \citet{cle83} and \citet{rob83}. This flare reached the peak  first at the higher frequencies and then propagated to millimeter wavelengths. In the attempt to explain this behaviour,  \citet{mar85} proposed a model in which a shock wave propagates in the relativistic jet and evolves though three different phases of energy loss: Compton, synchrotron and adiabatic. The time delay between the peak at different frequencies is a consequence of opacity, since the shock becomes  optically thin at the lower frequencies as the component expands. In this model, the X-ray and infrared flares should precede the radio flare with timescale of  months and in fact, the counterpart at 22 GHz of the infrared 1983 flare probably occurred 290 days later as proposed by \citet{bot88}.

\citet{ste94} detected typical time delays  of about 100 days between flares at $37$ and $90$ GHz in 17 sources monitored by \citet{all85} and in 3C273, \citet{ste98} reported time delays between the frequencies of 375  and 4.8 GHz. In all these situations, the time delay can be explained by the shocked jet models \citep{mar85,hu85} and their generalizations \citep{mar90,mar92,ste96,tur00,sok04}. Time delays even higher were predict by \citet{tur00} between radio and infrared wavelengths, with the flare at high energy always coming first. 

Before Fermi Gamma-ray Space Telescope started to operate, there was no much information about variability behaviour at $\gamma$-rays. EGRET, on board of GRO detected $\gamma$ -rays of energies above 100 MeV with an average value between 1991 and 1995 of $(1.5 \pm 0.2)\times 10^{-7}$ photons cm$^{-2}$s$^{-1}$, while \citet{col00} reported the largest flare observed with this instrument with a duration at least of 30 days and a flux density of $(7.6 \pm 0.2)\times 10^{-7}$ photons cm$^{-2}$s$^{-1}$.

 \citet{cha12} analysed the variability of six sources at the infrared, including 3C273, and found a positive correlation with the flares detected by Fermi/LAT at $\gamma$-rays with an upper limit of three days for the time delays. Even when a delay is detected, as in 3C279, it was  not  higher than a few days \citep{hay12}. Considering the short time delays between  the flares at these wavelengths, we interpreted that the 7 mm flares detected in 2010 March as the radio counterpart of the $\gamma$-ray flares detected by Fermi in 2009 September, in the same way as other radio flares were interpreted as radio counterparts, delayed  by several months, of  infrared flares \citep{bot88,ste94,ste98,tur00}.

As discussed in Section 2, VLBI observations reported  by \citet{jor12} show the appearance of four new superluminal components, ejected from the core between 2009.5 and 2010.5, at epochs coincident with the occurrence of the strong $\gamma$-ray flares. These coincidences also favours the shock model.  However the high intensity of the $\gamma$-ray flares compared to the moderate intensity of the radio flares still needs elucidation; the change of the beaming factor as a consequence of jet precession, as proposed by \citet{abr99} seems to be a good explanation, as discussed below. 

\begin{figure}
  \includegraphics[width=\columnwidth]{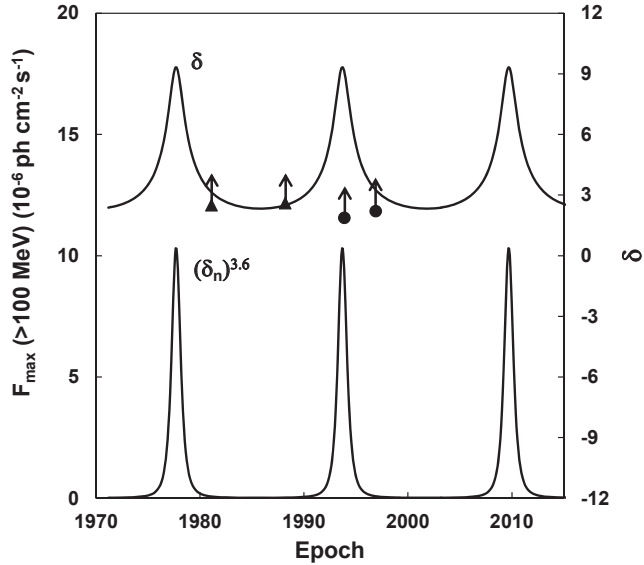}
  \caption{Doppler factor $\delta$ and beaming factor  of the $\gamma$-ray emission $(\delta_{\rm n})^{3.6}$, as a function of time, according to a 16 y precessing jet model, described in the text. The beaming factor $(\delta_{\rm n})$ was normalized  to match the  amplitude of the largest $\gamma$-ray flare. Full triangles and circle are lower limits imposed to  $\delta$ by X- and $\gamma$-ray observations, as described in Abraham \& Romero (1999). }
  \label{model}
\end{figure}

Both the radio and $\gamma$-ray emission must be boosted if the angle between the emitting region and the line of sight is small, but the effect in the flux density $S(\nu, \delta)$ is different at different frequencies, 
because in the observer reference frame: 

\begin{equation}
S(\nu, \delta)\propto \delta^{(p+\alpha)}\nu^{-\alpha}, 
\end{equation}

\noindent
where $\delta$
is the Doppler factor:

\begin{equation}
\delta = \frac{1}{\Gamma(1-\beta \cos\theta)},
\end{equation}

\noindent
and $\Gamma$ the Lorentz factor: 

\begin{equation}
\Gamma=\frac{1}{(1-\beta^2)^{1/2}},
\end{equation}

\noindent
$\beta$  is the bulk jet velocity, $\theta$ the angle between the jet direction and the line of sight, $\alpha$ the spectral index and, $p=2$ for a continuous jet and $p=3$ for discrete components.

For 3C273, the spectral index in the $\gamma$-ray region of the spectrum is $\alpha > 1$, while in the radio region, for the newly formed optically thick components until it becomes optically thin $\alpha \sim -2$. Therefore, an increase in the Doppler factor will have a strong effect in the flare emission at $\gamma$-rays, while would barely  affect the observed radio flux density. An increase in $\delta$ by a factor 3 or 4 seems to be a promising explanation for the relative intensities of the flares at these two frequencies.

\begin{figure}
  \includegraphics[width=\columnwidth]{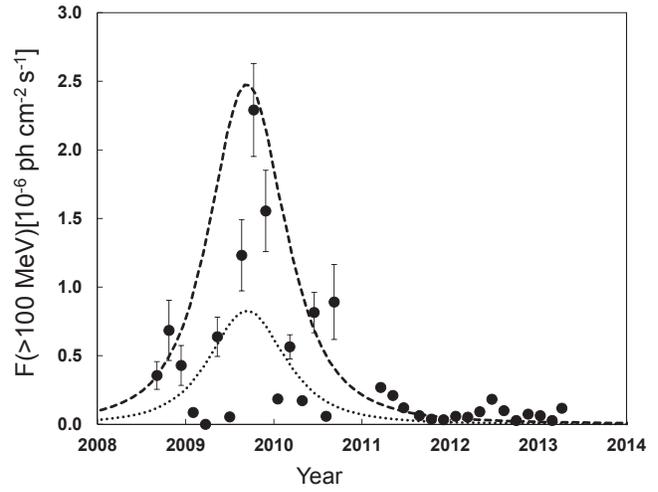}
  \caption{Fermi $\gamma$-ray light curve binned in 50 days (points) after discounting the minimum flux value from the data, which is of the order of  the upper detection  limit of the EGRET observations \citep{har99}. The dashed line  represents the flux of an arbitrary average flare  at different epochs due to beaming,  the dot line represents the same for a flare with an intrinsic intensity equal to $1/3$ of the average value.}
  \label{lc_model}
\end{figure}

Periodic changes in the Doppler factor of 3C273 were predicted by \citet{abr99}, as the result of changes in the angle between the jet direction and the line of sight, due to precession. The model was based on differences in the apparent superluminal velocities of components ejected at different epochs and in their position angles in the plane of the sky, and resulted in a precession  period of 16 years. During each cycle the model predicts a variation of a factor of three in $\delta$, and the  maximum approach between the jet and the line of sight to occur in 2010, when the boosting was maximum. Minimum values for the Doppler factor were obtained using limits imposed by early  X- and $\gamma$-ray observations \citep{abr99,col00}. We therefore calculated the beaming factor for the $\gamma$-ray photon flux density $\delta ^{p+\alpha+1}$ as a function of time, using $p=2$ and $\alpha + 1 = 1.6$ \citep{abd10}, and the values of $\delta$ as a function of time obtained from a precession model with a period of 16 years, an opening angle of the precession cone of $4\fdg 1$, an angle between the cone axis and the line of sight of $10^\circ$ and a Lorentz factor $\Gamma = 13$. This model fits the superluminal velocities of the jet components identified by \citet{abr96,lis09,jor12}. The Doppler and normalized beaming factor are presented in Fig. \ref{model}.

To compare the intensity of the $\gamma$-ray flares at different epochs we binned the $\gamma$-ray flux in intervals of 50 days, to obtain an error comparable to that of the light curve presented by \citet{col00}; the large time bin also eliminates the small timescale variations. The results are presented in Fig. \ref{lc_model} where the dashed line  represents the flux that an arbitrary average flare will have at different epochs due to beaming, and the dot line represents the same for a flare with an intrinsic intensity equal to $1/3$ of the average.  The low points values between 2009 and 2011 show the epochs with no flares.

Changes in the Doppler factor have other observable consequences. The first is the reduction in the time scale at the observer reference frame by a factor  $\delta / \delta_{\rm ref}$, where $\delta_{\rm ref}$ is the Doppler factor at a given reference time. A reduction in the time scale would imply in the increase in the ejection rate of superluminal components, and in fact, Jorstad et al. (2012) reported the ejection of four components coincident with the strong $\gamma$-ray flares, while the average rate of ejection is 0.7 y$^{-1}$, as can been seen in \citet{lis09b}. 

\begin{figure}
  \includegraphics[width=\columnwidth]{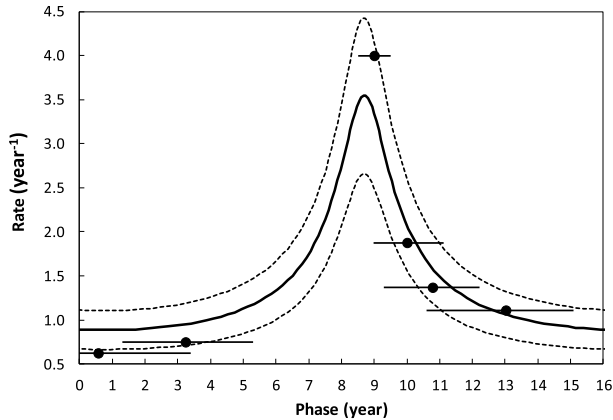}
  \caption{Ejection rate obtained from de precession model of components during the 16 year period. The dashed line represents a fluctuation of $25 \%$ in the rate, while the dots represents the observed rate. The phase 0 of the cycle was 1985 and the error bars represents the epoch used to determined the rate.}
  \label{rate}
\end{figure}

In Fig \ref{rate} we show the ejection rate variability as a consequence of the time contraction during one  precession cycle. Naturally, the intrinsic rate is not exactly the same every year and  the fluctuations are also amplify by the Doppler factor during its maxima value. In the figure, the dashed lines represent  fluctuation of $25 \%$  in the intrinsic rate at the source frame. It is not easy compare the predicted rate with the observational data because the ejection time of the components is not an observed quantity, and it is necessary to compute the component kinematics. The observed ejection rates were obtained using different works, they are presented in table 2 and shown by dots in  Fig. \ref{rate}, with error bars that indicate the interval at which the rate was obtained.  
We only used time intervals smaller than five years, and those in which the components were close to the core to guarantee that short lived components were not missed.           

\begin{table} 
 {\scriptsize
 \caption{Ejection Rate for superluminal components}
 \hfill{} 
 \begin{tabular}{l l l l}
\hline number of & rate  &  epoch & reference\\
components &  (year$^{-1}$)& & \\
\hline 4 & 4.0 & 2009.5-2010.5 & \citet{jor12}\\
4 & 1.4 & 1994.3-1997.2 & \citet{lis09}\\
4 & 1.9 &  1996.1-1994 & \citet{hom01}\\
5 & 0.8 & 1979.6-1984.1 & \citet{tur99}\\
3 & 1.1 & 1986.3-1990.3 & \citet{tur99}\\
3 & 0.6 & 1983.6-1988.4 & \citet{abr96}\\
\hline
\end{tabular}}
\hfill{}
\label{T2}
\end{table} 

The second consequence of a change in $\delta$ is its effect on the time delay $\tau$ between the radio and $\gamma$-ray flares. According to the shock model of Marscher \& Gear (1985): 

\begin{equation}
\tau(\delta, \nu)=\frac{R'(\nu')(1+z)}{\Gamma\delta c},
\end{equation}

\noindent
where $R'$ is the distance to the origin of the shock at which the source becomes optically thin at the frequency $\nu'$, both measured 
in the source reference frame, $z$ is the redshift and $c$ the speed of light.
If the source is already in the adiabatic energy lose phase:

\begin{equation}
 R'(\nu')\propto (\nu')^{-\omega}
\end{equation}

\noindent
where $\omega=(3s+12)/(7s+8)$, and  $s=1-2\alpha$ is the index of the electron energy distribution.

Since $\nu'=\nu/\delta$, we obtain from equations (4) and (5):

\begin{equation}
\tau(\delta, \nu)=\tau(\delta_{\rm ref}, \nu)\left(\frac{\delta}{\delta_{\rm ref}}\right)^{(\omega-1)}
\end{equation}

In the optically thin regime $\alpha \sim -0.7$, resulting in  $\omega=0.77$. For $\delta/\delta_{\rm ref}$ equal to 3 or 4, the ratio  $\tau(\delta, \nu)/\tau(\delta_{\rm ref}, \nu)$ will be 0.78 and 0.73, respectively.

To evaluate the expected time delay $\tau(\delta,\nu)$ between the 7 mm and $\gamma$-ray flare, we use as a reference the flare that occurred in August 1995, and was detected at wavelengths ranging from 0.8 mm to 6.25 cm \citep{ste98}. Since there are no $\gamma$-ray observation of this event, we assume that the maximum at 0.8 mm coincided with the origin of the flare, resulting in a time delay at 7 mm  of 234 days, which gives $\tau(\delta, \nu)$ = 182 or 170 days for $\delta / \delta_{\rm ref}$ equal to 3 or 4, respectively, which agrees very well with the observed delay.

\subsection{The precession model and the historic light curve}

Since the radio variability  is mainly due to the outbursts produced by the appearance of new  jet components, the radio light curve should not be too much affected by the periodic Doppler variation, as discussed in the previous subsection.  Furthermore, the formation of  new components is not necessarily periodic, and in fact, the historic light curve compiled by \citet{sol08} shows maxima and minima that do not seem to follow a periodic pattern, instead of that, they occur with different intensities and time intervals.  However, the ejection rate of these components and their temporal evolution  are dependent of the Doppler factor and can introduce a modulation in the radio light curve. 

To verify the existence of periods, we used two different statistical test: Stellingwerf and Structure Function. Unfortunately, to detect a  16 year periodicity it is necessary a long time coverage for the light curve; for example, in 20 year interval only one cycle was completed, which turns the 16 year periodicity detection impossible. However, 3C273 is one of the best monitored radio sources in the sky,  and the historic data covers 37 years at some frequencies \citep{sol08}. The Stellingwerf method is not adequate to detected periodicities higher than $1/5$ of the total time coverage, and it could not reveal the 16 year period without 80 years of radio observations. However, it can be very useful to identify possible resonant periodicities, as the 8 year periodicity detected in many works with other methods \citep{fan07,vol13}.

\begin{figure}
  \includegraphics[width=\columnwidth]{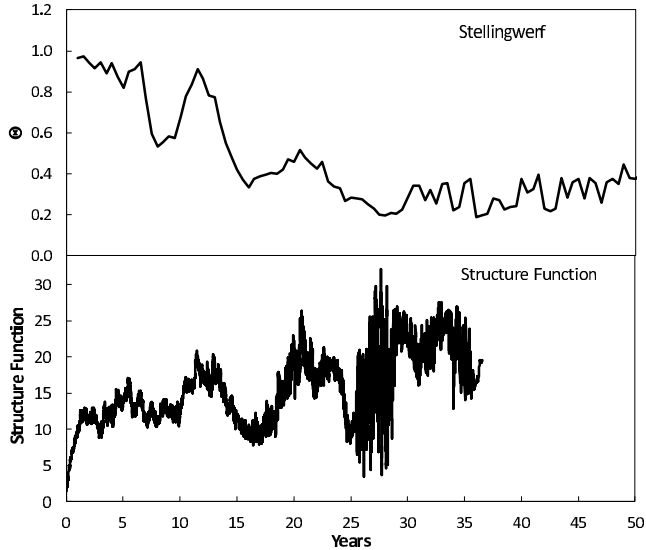}
  \caption{Upper part: Stellingwerf method result, which indicate a periodicity of $8$ years. Lower part: Structure Function results, which indicate a periodicity of 16 years.}
    \label{Periods}
\end{figure}

The Stellingwerf method \citep{ste78} divides the sample in $m$ groups, following a phase vector given by:
\begin{equation}
 \phi_{i}=\frac{t_{i}}{P}-\biggl [\frac{t_{i}}{P} \biggr ],
\end{equation}
\noindent 
where $P$ is the guessed periodicity, $t_{i}$ is the time of observation and the bracket means the integer part. 
For a given $P$, the Square Deviation of the flux density in each group is computed  and the sum of all Square Deviations is divided by the total Square Deviation  of the sample. The result of this fraction is defined as $\theta$, and it will be minimum when the guessed periodicity  is the real one. The minimum in the $\theta$ versus period curve will be deeper when $m$ is higher, however, $m$ can not be excessively high to allow a reliable statistics for each group; in our work we choose $m = 20$. 

The Structure Function  (SF) is a simply way to verify how much the intensity varies after a given time $\xi$ and  is given by: 
\begin{equation}
SF(\xi)= \langle [I(t+\xi)-I(t)]^{2} \rangle,
\end{equation}
\noindent
where $I$ is the intensity at time $t$ and $I(t+\xi)$ is the intensity after a time delay $\xi$. If there is indeed a periodicity $T$, when for $\xi = T$, there will be a minima in the $SF$ curve.

Considering the different wavelengths presented in the historic data reported by \citet{sol08}, we choose the $37$ GHz light curve to discuss the results of both methods, because it is the nearest frequency to our observations. However, we performed the tests at the other radio frequencies and the result are very similar, as already noted by \citet{vol13}. The 37 GHz data started at 1970 and ended at 2006, which gives a time coverage of almost $37$ years; it is one of the frequencies with more observations days.

 The result of the statistical tests are present in Fig. \ref{Periods}, the Stellingwerf method in the upper part and Structure Function in the lower part.
The result of the Stellingwerf method shows a prominent minimum at 8 years, and a beginning of an even deeper minimum  at 16, before  the appearence of oscillations that are the consequence of insufficient time coverage. The significance of the minima is obtained applying the F-test, where $f=(1-\Theta)/\Theta$, needs to be higher than 0.25 \citep{kid92}, condition satisfied for the 8 year minimum, where $f=0.86$.  We interpreted this detection, also found  \citet{vol13} using  Fourier Analysis, as a resonance of the 16 years precession period.
The 16 year minimum also presents a high $f$ value, however, as pointed out above, only the first period is inside the limit of what can be detected by the Stellingwerf method. 

\begin{figure}
  \includegraphics[width=\columnwidth]{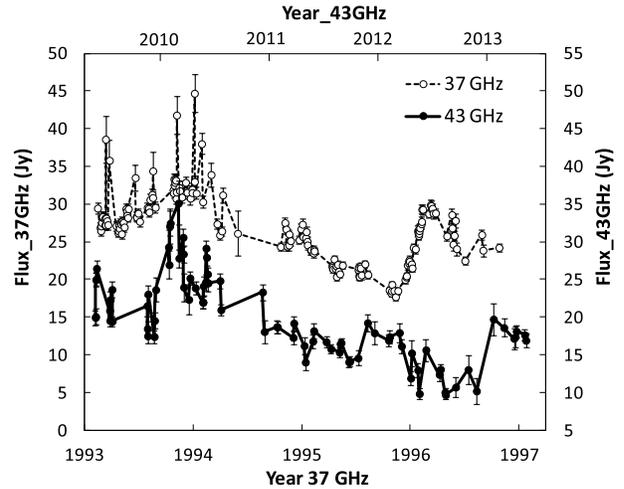}
  \caption{37 GHz variability obtained between 1993 and 1997 \citep{sol08} and the 43 GHz variability obtained in this work, between 2009 and 2013. Both light curve show the same variability behaviour.}
  \label{lc2}
\end{figure}
The SF intensity variation does not show any minima around 8 years, but shows a wide minimum around 16 years. The large width, which represents an imprecision in the detection of about 2 years, can be attributed  to the small time coverage and to the fact that the formation of new components in the jet is not a strictly periodic phenomena.  The result of both tests are consistent with the 16 years precession model, because the Stellingwerf method is efficient to detected the  8 year resonance while the SF revels only the 16 year period. 

Based on the statistical results, we compared the $43$ GHz light curve obtained between 2009 and 2013, with that at $37$ GHz, obtained 16 years earlier \citep{sol08}, as shown in Fig. \ref{lc2}. Besides the $7$ Jy difference in flux density, due to the difference in frequencies, both light curves show the same variability pattern, with the same long decrease in flux density that lasted for two years. Again, the individual  oscillations do not match very well, as expected  from the fact that the radio outbursts are not strictly periodic.

\section{Conclusion}
\label{conclusion}

We presented the results of four years monitoring of 3C273 at 7 mm. During this period we detected a flare in 2010 March, that we interpreted as the radio counterpart of the extremely intense $\gamma$-ray flare observed by Fermi/LAT in 2009 September delayed by approximately 170 days.  This delay can be understood  in the context of the shock model in which the electrons are accelerated in a shock that propagates along the jet, originating the $\gamma$-ray flare though the Inverse Compton process and later the radio flare, when the shock turns optically thin at this lower frequency. 

We explained the very high intensity of the $\gamma$-ray flare compared to previous ones as the consequence of boosting, produced by an increase in the Doppler factor by  3 or 4. The intensity of the radio emission, on the other hand, would not be affected, because the source was optically thick at these frequencies. A periodic variation of the Doppler factor  was predicted by the precession model of \citet{abr99} for the jet of 3C273. The precessing period was 16 years and the parameters of the precessing jet used in the present work were: opening angle of the precession cone of $4\fdg 1$,  angle between the cone axis and the line of sight of $10^\circ$ and  Lorentz factor $\Gamma = 13$. 

Other observable consequences of the variation of the Doppler factor are the increase in the rate of superluminal ejections, which was confirmed by the work of  \citet{jor12}, and its effect on the time delay between flares at different frequencies, which was also compatible with the observations.
 
Although  the Doppler factor does not affect the radio flux density, it modulates the radio light curve, as a consequence of the difference in the ejection rate of jet components and their temporal evolution. The Stellingwerf method and the Structure Function, calculated from the $37$ GHz   historic light curve, covering almost 40 years,  revealed the existence of 8 and 16 year periodicity. We interpreted the first one, only detected by the Stellingwerf method, as a resonance of the 16 year period. Moreover, the variability pattern detected between 2009 and 2013 has the same trend than that detected 16 years earlier at 37 GHz.

\section*{Acknowledgments}
We are grateful to the Brazilian research agencies FAPESP and CNPq for financial support. This study makes use of 43 GHz VLBA data from the Boston University gamma-ray blazar monitoring program
(http://www.bu.edu/blazars/VLBAproject.html), funded by NASA through the Fermi Guest Investigator Program. This research has made use of data from the MOJAVE database that is maintained by the MOJAVE team (Lister et al., 2009, AJ, 137, 3718).

\end{document}